\documentclass[%
 pre,
 amsmath,amssymb,
 reprint,%
]{revtex4-1}

\usepackage{graphicx}
\usepackage{dcolumn}
\usepackage{bm}

\usepackage{xcolor}
\usepackage{color,soul}
\usepackage{hyperref}
\usepackage{colortbl}
\usepackage{lipsum} 

\usepackage{geometry}
\newcommand{\beq}{\begin{equation}}
\newcommand{\eeq}{\end{equation}}
\newcommand{\cs}{\mathrm{cs}}

\begin{document}



\title{Autoresonant laser acceleration of electrons in a strongly magnetized plasma solenoid}

\author{{Iu} Gagarin}
\author{Ph Korneev}
\email{ph.korneev@gmail.com}
\affiliation{P.N. Lebedev Physical Institute of the Russian Academy of Sciences}

\date{\today}

\begin{abstract}
Direct laser acceleration of electrons is considered in a strongly magnetized plasmoid with the magnetic field strength allowing for reaching the auto-resonance. The  plasmoid may be optically created by irradiation of specially designed targets with an auxiliary intense laser beam at the previous stage of interaction in a possible all-optical setup. Specifics of the strongly magnetized plasma solenoid may be critically important for the resonant processes where a small deviation of the parameters destroys the matching conditions. The process of the autoresonant electron acceleration is analyzed for different configurations inherent in the possible realizations of the setup, estimates for the efficiency of acceleration and resonance magnetic fields are proposed. Despite the modifications of the autoresonance conditions and the presence of collective effects, generation of energetic electron bunches in a realistic setup may be possible and effective.
\end{abstract}

\maketitle

\section{\label{intro}Introduction}

A tremendous progress in laser techniques in a few past decades allowed for routine generation of short relativistic pulses, which eventually became the most perspective acceleration setups, at least for electrons. Different discussed laser acceleration schemes (DLA \cite{dla}, LWFA \cite{lwfa_1, lwfa_2}, etc.) show efficient electron acceleration producing GeV pC \cite{gev_1, gev_2} and MeV $\mu$C \cite{mev_muc} bunches which can be used e.g. as x-ray sources. In this context, autoresonance laser acceleration (ALA) may be considered as another promising scheme \cite{cara} with multiple advantages, for example, a high energy gain on small acceleration length, a small angular divergence of the accelerated electron bunch, an opportunity to generate ultrarelativistic electrons in near-resonant regime \cite{plane_analytic}. On the other hand, matching the resonant conditions requires either pre-accelerated to 10-100 MeV \cite{sin_env_res} electrons or strong magnetic fields of the order of $\sim 10$ kT, which makes the application of ALA questionable.

However, recent theoretical findings related to the optical quasi-stationary magnetic field generation of multi-kT scale \cite{Zosa.etal_100kTMagneticField_APL-2022,Bukharskii.Korneev_StudyHighlyMagnetized_BLPI-2023,Longman.Fedosejevs_KiloTeslaAxialMagnetic_PRR-2021,Liseykina2016,chiral,Dmitriev.Korneev_LaserPulseInteraction_BLPI-2023} and a certain experimental progress in this direction \cite{Santos.etal_LaserdrivenPlatformGeneration_NJP-2015,Santos.etal_LaserdrivenStrongMagnetostatic_PoP-2018,Law.etal_RelativisticMagneticReconnection_PRE-2020, Kochetkov.etal_NeuralNetworkAnalysis_SR-2022} make the realization of all-optical ALA much closer. Especially interesting in this context is a fiber setup with a transverse internal structure, directing dischargesinduced by laser heating \cite{chiral, Dmitriev.Korneev_LaserPulseInteraction_BLPI-2023}. In this setup, the laser pulse may almost fully be absorbed and create quite a long strongly magnetized plasma channel. In this case, it becomes possible to achieve the autoresonance regime of acceleration with initially cold electrons, i.e. no electron pre-acceleartion is needed.

{Theoretical study of electron acceleration by plane waves in a uniform magnetic field was presented in} \cite{kolom}. It was shown that in isotropic refractive medium the resonance condition can be maintained automatically for the entire motion when the refractive index $n\to1$. 
Other effect on the autoresonance is related to the radiation reaction force, which leads to a slow dephasing and an eventual resonance breaking \cite{rad_react}.
Analytical expressions for accelerates electron coordinates, momenta and energy, in general case for the plane electromagnetic wave with arbitrary polarization can be derived as functions of the wave phase 
\cite{gen_pol+spectrum}.

Other configurations of charged particle acceleration schemes with plane electromagnetic waves, for example, in resonators or with appropriate magnetic field spatial profile have been considered, showing a wide variety of possible applications of the autoresonance mechanism \cite{review}.
Some schemes suggest utilizing {powerful} terahertz electromagnetic pulses \cite{terahertz_acc}, which becomes interesting in the context of extensive studies  of controlled THz radiation generation in recent years \cite{thz_1, thz_2, Bukharskii.Korneev_IntenseWidelyControlled_MRE-2023, Dmitriev.etal_PowerfulEllipticallyPolarized_P-2023}. Due to the lower frequency of the accelerating wave in this case, the resonance condition can be met with more feasible longitudinal magnetic fields of smaller magnitudes.


 \begin{figure}[]
     \centering
     \includegraphics[width=\linewidth]{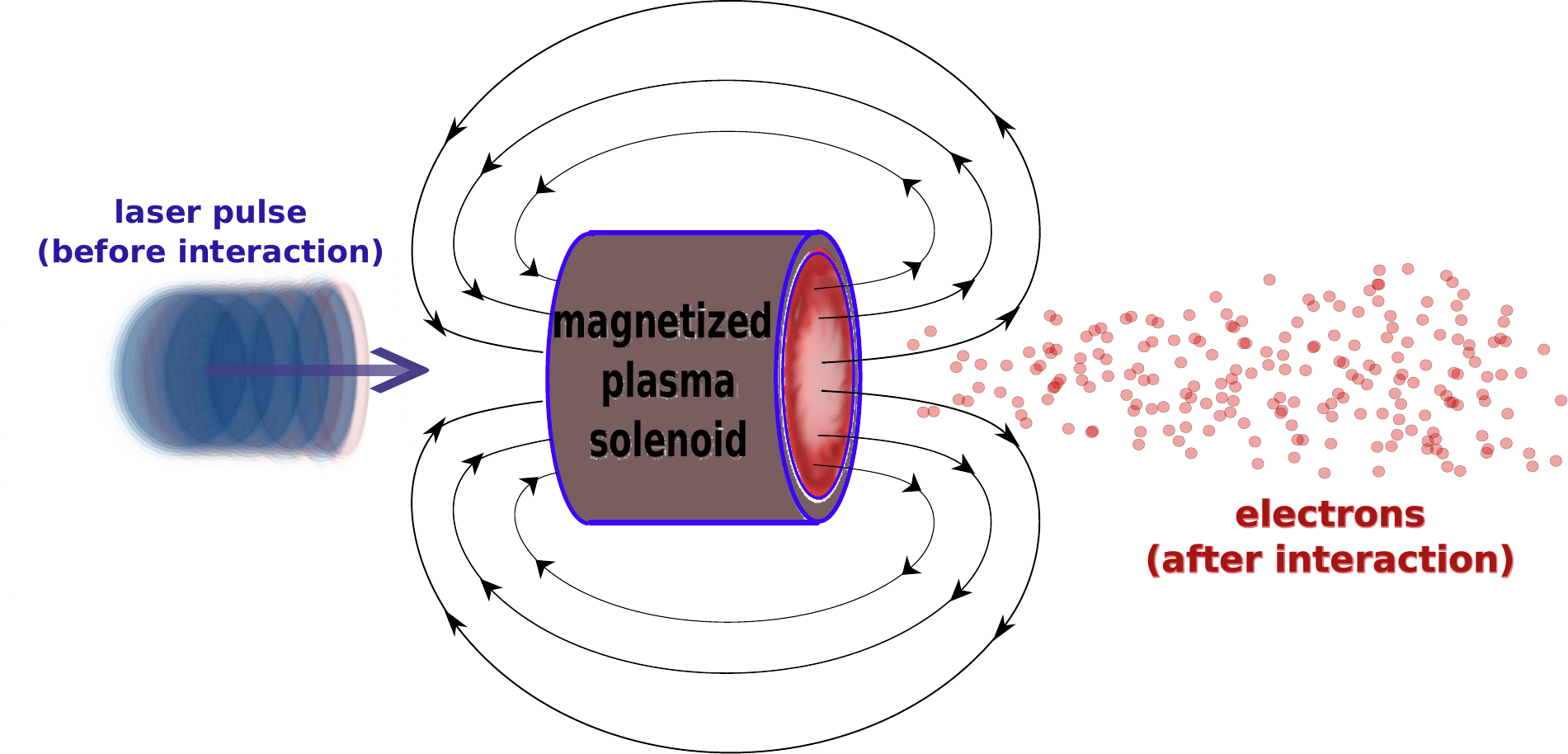}
     \caption{A possible interaction scheme. An intense laser beam is focused along the magnetic field lines in plasma to enable autoresonant acceleration process.}
     \label{fig:scheme}
 \end{figure}

In case of an experimental realization of the autoresonant condition in infra-red or visible light range without pre-acceleration of electrons, the irradiated system may actually be performed as a strongly magnetized plasmoid, as the magnetic fields of the required several kT may be probably supported only in hot plasmas \cite{Ehret.etal_KiloteslaPlasmoidFormation_PRE-2022}. In such kind of systems, the ALA looks feasible, however there are questions related to its robustness in terms of the combined effects of the magnetic field shape, plasma effects, and the laser pulse profile. It is expected, that e.g. intrinsic magnetic field inhomogeneities in any real experiment may significantly change electron dynamics and decrease the effectiveness of the acceleration scheme. 

This work is aiming at studies of the influence of realistic experimental nonidealities on the autoresonant acceleration in plasma micro-structures with ultra-strong magnetic fields.
The paper is organised as follows: after the introduction, analytical description of the ALA is reminded and estimates for the resonant magnetic field are presented in Section \ref{analytic}. The analysis further is continued numerically in Section \ref{numerics}. Then after summing up the obtained results, the conclusions are presented.

\section{\label{analytic}Analytical estimates for the autoresonant acceleration conditions}

The seminal works related to the autoresonant charged particle acceleration \cite{rob_bush,kolom,davydov} revealed several key features of the process. In the auto-phase matching condition between the particle velocity and the laser field a crucial role plays the integral of motion of the accelerated particle, known also as a "dephasing rate" 
\beq
\label{Gamma}
\Gamma = \gamma - \frac{P_x}{mc},
\eeq where $\gamma=\sqrt{1+\mathbf{P}^2/(mc)^2}$ is the relativistic factor, $\mathbf{P}$ and $P_x$ are the particle momentum and its longitudinal projection, $m$ is the particle mass, hereafter the particle means electron \footnote{Other charged particles with larger masses require even stronger magnetic fields to match the resonance with the optical frequencies.}, and $c$ is the light velocity. So, if the resonance condition $\omega_c = \Gamma \omega$, where $\omega$ is the laser frequency, $\omega_c=e B_0 /(mc\gamma)$ is the cyclotron frequency for the magnetic field with the amplitude $B_0$, $e$ is the elementary charge, is satisfied initially, then it is maintained during the particle acceleration process due to the conservation of $\Gamma$. Physically it means that electron rotates in the longitudinal magnetic field in phase with the laser electric field thus continuously gaining energy. 
An equivalent explanation of the autoresonance nature of the acceleration process is that the charged particle spectrum in a homogeneous magnetic field (Landau levels) is equidistant \cite{equidistant}.  

The ideal case of the plane wave should be reconsidered for experimentally feasible situation of intense laser pulses. Gaussian profile provides a more realistic description of electromagnetic field spatial distribution, though complicates the equations. For the autoresonance electron acceleration in such pulses the integral of motion $\Gamma$ is not conserved, and automatic phase matching no longer takes place. 
For an ensemble of particles, the resonant conditions may be violated for some of them, e.g. due to Coulomb collisions, which lead to small deviations in electron energy distribution \cite{coulomb}, or due to the radiation reaction force \cite{rad_react,plane_analytic}. Also the magnetic field profile affects the achievable parameters of the accelerated electron beams.

Remind first the main equations and the required autoresonant conditions. This would help to understand how the laser pulse and the magnetic field shapes, as well as plasma effects may change the frequency matching and how the acceleration in experiment may change.

\subsection{The autoresonant condition in a plane wave}

Consider a plane linearly polarized wave with a $\sin$-square envelope, propagating in the positive direction of the $X$ axis, and a constant magnetic field $\mathbf{B}_0$ directed along the laser pulse propagation. The fields may be written as 
\begin{eqnarray}
    \label{E}
    \mathbf{E} & = & a_0 \frac{mc\omega}{e}\mathbf{e_y} ~\sin^2\frac{\pi\xi}{\omega T} ~\cos\xi , \\
    \mathbf{B} & = & a_0 \frac{mc\omega}{e}\mathbf{e_z} ~\sin^2\frac{\pi\xi}{\omega T} ~\cos\xi  + b_0 \frac{mc\omega}{e} \mathbf{e_x}, \label{B} 
\end{eqnarray}
where $\xi  =  \omega \left(t - {X}/{c}\right)$ is the wave phase, $a_0=eE_0/(mc\omega)$ and $b_0=eB_0/(mc\omega)$ are the dimensionless amplitudes of the laser field with the amplitude $E_0$ and the applied constant magnetic field respectively, $T$ is the laser pulse temporal length. In these notations, the cyclotron frequency may be rewritten as 
\beq
\omega_c={b_0 \omega}/{\gamma}.
\label{cyclotron}
\eeq

Equations of motion with an equation for energy governs electron dynamics:
\begin{eqnarray}
    \frac{d \mathbf{R}}{dt} = \frac{\mathbf{P}}{m \gamma}, \label{eqm1}\\
    \frac{d \mathbf{P}}{dt} = -e \left(\mathbf{E} + \left[ \frac{\mathbf{P}}{mc \gamma} \times \mathbf{B} \right]\right), \label{eqm2}\\
    m^2 c^2 \gamma \frac{d \gamma}{dt} = -e (\mathbf{P} \cdot \mathbf{E}).\label{eqm3}
\end{eqnarray}

Approximate solutions of this system as functions of laboratory time are given in \cite{kolom,rob_bush}. Another approach considering electron properties as functions of the phase $\xi$ leads to exact solutions of the system \cite{plane_analytic}. Crucial role in solving equations of motion plays the integral \eqref{Gamma} as was mentioned already in Refs. \cite{davydov,kolom,rob_bush}. It can be obtained directly from the equations of motion, by projecting of \eqref{eqm2} on $\mathbf{e}_x$ and expressing $(\mathbf{P}\cdot\mathbf{E})$ from \eqref{eqm3}. 
Time derivative of the phase $\xi$ considered on the electron trajectory simplifies with the use of $\Gamma$:
\begin{eqnarray}
    \label{xi}
    \frac{1}{\omega} \frac{d\xi}{dt} = \frac{1}{\gamma} \left(\gamma - \frac{P_x}{mc}\right) \equiv \frac{\Gamma}{\gamma}
\end{eqnarray}

For the plane wave $E_y=B_z$ with an appropriate choice of dimensionless variables $\mathbf{r}=(x,y,z)={\omega}\mathbf{R}/c=(\omega X/c,\omega Y/c,\omega Z/c)$, $\mathbf{p}=(p_x,p_y,p_z)={\mathbf{P}}/{mc}=(P_x/mc,P_y/mc,P_z/mc)$ and using \eqref{xi} obtain the following {system of equations}:
\begin{eqnarray}
    &&\frac{d\mathbf{r}}{d\xi} = \frac{\mathbf{p}}{\Gamma}, \label{sys1}\\
    &&\frac{d p_x}{d\xi} = - \frac{ a_0}{\Gamma}~ p_y~\cs(\xi,T), \label{sys2}\\
    &&\frac{d p_y}{d\xi} = - a_0~\cs(\xi,T) - \frac{b_0}{\Gamma}~ p_z , \label{sys3} \\
   && \frac{d p_z}{d\xi} = \frac{b_0}{\Gamma}~ p_y  \label{sys4},
\end{eqnarray}
where $~\cs(\xi,T)\equiv\sin^2\frac{\pi\xi}{\omega T} ~\cos\xi$ is the temporal dependence of the laser wave. By derivation \eqref{sys3} and using \eqref{sys4}, an equation for the driven harmonic oscillator is obtained:
\begin{eqnarray}
    \frac{d^2 p_z}{d \xi^2} + \left(\frac{b_0}{\Gamma}\right)^2 p_z = -\frac{a_0 b_0}{\Gamma} ~\cs(\xi,T)
\end{eqnarray}

To define explicitly the resonant conditions, rewrite $\cs(\xi,T)$ in the right-hand side as
\begin{equation}
    \cs(\xi,T)= \frac{1}{4}\left(2\cos(\xi) - \cos(\omega_+ \xi) - \cos(\omega_- \xi)\right), 
\end{equation}
where 
\beq
    \omega_{\pm} = 1\pm \frac{2\pi}{\omega T}.
\eeq
Therefore the particle in the plane wave and the longitudinal magnetic field gain energy when the magnetic field is strong enough
\beq
    \frac{b_0}{\Gamma} = \{1, \omega_{\pm}\},
\eeq
or 
\beq
    \frac{\omega_{ce}}{\gamma} = \omega \left(1 - \frac{v_x}{c}\right), \\
    \frac{\omega_{ce}}{\gamma} = \left(\omega \pm \frac{2\pi}{T}\right) \left(1 - \frac{v_x}{c}\right).
\eeq
Note that for initially unmoving partcile $\Gamma = 1$ and $\gamma = 1$, so the cyclotron frequency should be equal (for $b_0=1$) or close (for $b_0=\omega_{\pm}$) to the laser frequency, which corresponds to $B_0\sim 10$kT for the optical range.

For the plane wave, the integral of motion $\Gamma$ provides an automatic frequencies matching if the resonance condition is matched at the initial time moment
\begin{equation}
    \omega \left(1 - \frac{v_x}{c}\right) = \omega_c,
    \label{resonance}
\end{equation}
this is why with moderate magnetic fields, an initial pre-acceleration of the particles may help to reach the resonance condition.

Spectrum radiated by accelerated electrons is numerically studied in \cite{sin_env,sin_env_res} (with sin-square envelope) and \cite{plane_analytic} (without an envelope). It is shown that in the direction along the magnetic field there are two main peaks: on the laser and cyclotron frequencies. Sin-square envelope leads to the generation of two additional peaks in the radiated spectrum symmetrically sided near the cyclotron frequency.

\begin{widetext}
For the considered case of the plane wave, assuming a very slow-varying envelope, the solutions for the electron momentum read
\beq
     p_z = C_0 \left[\cos\left(\frac{b_0 \xi}{\Gamma}\right) - \cos\left(\xi\right)\right]
    - C_+ \left[\cos\left(\frac{b_0 \xi}{\Gamma} \right) - \cos\left(\omega_+ \xi\right)\right]
    - C_- \left[\cos\left(\frac{b_0 \xi}{\Gamma} \right) - \cos\left(\omega_- \xi\right)\right], \label{exactp1}
\eeq
\begin{multline}
     p_y =  \frac{\Gamma}{ b_0} \left\{ -C_0 \left[\frac{b_0 }{\Gamma} \sin\left(\frac{b_0 \xi}{\Gamma}\right) 
    - \sin\left(\xi\right)\right] \right. + \label{exactp2}\\
    + C_+ \left[\frac{b_0}{\Gamma} \sin\left(\frac{b_0 \xi}{\Gamma}\right) - \omega_+ \sin\left(\omega_+ \xi\right)\right]
    \left. + C_- \left[\frac{b_0}{\Gamma} \sin\left(\frac{b_0 \xi}{\Gamma} \right) - \omega_- \sin\left(\omega_- \xi\right)\right] \right\},  
\end{multline}
\beq
      p_x = \frac{1 - \Gamma^2 +  p_y^2 +  p_z^2}{2\Gamma}, \label{exactp3} 
\eeq
where the constants are defined as 
\beq
     C_0 = \frac{{a_0 b_0 \Gamma}}{2 \left(b_0^2 - \Gamma^2\right)}, \quad C_{\pm} = \frac{ a_0 b_0 \Gamma}{4 \left(b_0^2 - \omega_{\pm}^2\Gamma^2\right)}. \label{exactp4}
\eeq
\end{widetext}
{The scaling for the electron momentum and energy at the exact resonance for $\xi \gg 1$ then reads}
\begin{eqnarray}
    \label{scalings_plane_wave}
    p_x \sim \xi^2, \qquad p_y \sim \xi, \qquad p_z \sim \xi.
\end{eqnarray}
Of course, the acceleration is limited due to the finiteness of the laser pulse even in the plane wave. However, in a real experimental situation, the geometry of the interaction and therefore the particle positions may be the determining factors. 

Considering ultrarelativistic particles the radiation reaction effect should be considered. In an external magnetic field the intensity of radiation can be written as \cite{landau}
\beq
    I = \frac{2 e^4}{3 m^2 c^3} \mathbf{B}^2 \gamma^2 = \frac{2 e^2 \omega^2}{3 c} \gamma^2 b_0^2.
\eeq
The radiation reaction force for an ultrarelativistic electron is estimated the in terms of the intensity as $F_{rr} \sim \frac{I}{c}$. The Lorentz force can be estimated as $F_L \sim a_0 m \omega c$. For the relation between the radiation reaction force and the Lorentz force in an ultrarelativistic limit one obtains
\beq
    \label{ratio}
    \frac{F_{rr}}{F_L} \sim \frac{2 b_0^2}{3 a_0} \gamma^2 \frac{e^2 \omega}{m c^3}.
\eeq

\subsection{The autoresonant condition in a Gaussian laser pulse}

Reaching high intensities and therefore high acceleration rates requires a sharp focusing of the laser beam, so that it becomes sufficiently different from a plane wave. For an estimate, in this subsection consider a Gaussian laser profile instead of the plane wave \eqref{E} and \eqref{B}. Although in further development different profiles may be found more interesting in the autoresonant acceleration context, the Gaussian profile adds to the problem new geometric parameters, such as a finite transverse beam waist and a spatial dependence of the intensity along the propagation direction. In the paraxial approximation laser electric and magnetic fields (without a slow temporal envelope) are given by (see, e.g. \cite{gaussian_pulse})
\beq
    \mathbf{E_L}  =  a_0 \frac{m c \omega}{e} \frac{w_0}{w}  ~e^{\displaystyle  -\frac{R^2}{w^2}} \cos(\xi - \psi_R + \psi_G) \mathbf{e_y}, \label{gaussE}
\eeq
\beq
    \mathbf{B_L}  =  a_0 \frac{m c \omega}{e} \frac{w_0}{w}  ~e^{\displaystyle  -\frac{R^2}{w^2}} \cos(\xi - \psi_R + \psi_G) \mathbf{e_z}, \label{gaussB}
\eeq
where
\beq
    w  =  w_0 \sqrt{1 + \frac{X^2}{X_r^2}}, ~ R^2 = Y^2 + Z^2,~ 
    \psi_R  =  \frac{k R^2}{2 R_0},
\eeq
$w_0$ is the beam waist radius at the focus, $X_r = {k w_0^2}/{2}$ is the Rayleigh length, $k = {\omega}/{c}$ is the wave number, $R_0 = X + {X_r^2}/{X}$ is the curvature of the wave front, $\psi_G = \arctan\left({X}/{X_r}\right)$ is the Gouy phase.

Resonance condition includes the derivative of the phase
\begin{eqnarray}
    \frac{d \Psi}{dt} \equiv \frac{d}{dt} \left(\xi - \psi_R + \psi_G\right) = \omega_c,
    \label{time_der_phase}
\end{eqnarray}
which was simply expressed with the constant integral of motion in the plane wave \eqref{Gamma}. For the Gaussian beam there are two changes: (i) the geometrical position of the particles affects the resonance due to the intensity profile, the resonance modifies, and the magnetic field should be defined from some average intensity level; (ii) the phase structure of the wave would destroy the resonance at some distance, so that the maximum acceleration depends not only on the pulse duration, but also on the Rayleigh length.

Time derivative of the phase $\xi$ leads to the resonance condition in the form \eqref{resonance}, while other phases give rise to correction terms. They can be estimated as:

\begin{equation}
    \dot \psi_G = v_x \frac{X_r}{X^2 + X_r^2} \leq \frac{v_x}{X_r} = \omega \frac{v_x}{c} \frac{2}{k^2 w_0^2}, 
\end{equation}
\begin{equation}
\dot \psi_R = \frac{k R}{2 R_0^2} (2 R_0 \dot R - R \dot R_0),
\end{equation}
and the rough estimate for the amplitude of the phase deviation in \eqref{time_der_phase} due to the pulse profile is then 
\begin{equation}
    \dot \psi_G - \dot \psi_R \lesssim \mathrm{C_{GR}} \frac{\omega}{k^2 w_0^2},
    \label{gauss_res_cor}
\end{equation}
where it was assumed that the electron is accelerated near the focus of the laser pulse, so that $R\lesssim w_0$, and $\mathrm{C_{GR}}\sim 1...10$. So, the focusing of the laser pulse would affect the resonant, reducing the overall effectivity of the acceleration. Besides, this estimate does not take into account the finite time of acceleration, which is defined by the time of passing the Rayleigh length by accelerating electron. 

\subsection{The role of an inhomogeneity of the magnetic field}

The role of the magnetic field inhomogeneity is analyzed with the transverse Gaussian profile
\begin{equation}
    \label{magn_field}
    \mathbf{B_{ex}} = B_0 \exp\left(- \frac{R^2}{2 \sigma^2}\right) \mathbf{e_x} 
\end{equation}

The resonance condition (\ref{resonance}) for the magnetic field averaged over the cross section bound with the characteristic electron radius leads to an amplitude dependency on the spatial scale:
\begin{eqnarray}
    \label{gauss_full}
    B_0 =  \frac{\Gamma ~E_0 ~l^2}{2 \sigma^2 (1 - \exp(-\frac{l^2}{2 \sigma^2}))}  , \\
    B_0 \approx \Gamma~ E_0 \left(1 + \frac{l^2}{4 \sigma^2}\right)  \quad \text{for} \quad \frac{l^2}{2 \sigma^2} \ll 1,
    \label{magn_param}
\end{eqnarray}
where $l$ is the characteristic electron transverse displacement.

This estimate can be used to calculate the shift in the resonant magnetic field and thus defines it in case of the Gaussian profile. New resonant magnetic field is the one leading to the highest energy gain in a plane wave pulse. In numerical simulations expressions \eqref{gauss_full}, \eqref{magn_param} provide an appropriate magnetic field amplitude $b_0$ for the given inhomogeneity spatial scale $\sigma$.

\subsection{Estimations for realistic parameters of the laser autoresonant acceleration}
Using electron momentum scalings \eqref{scalings_plane_wave} for realistic parameters of a typical laser experiment and a sufficiently strongly magnetized plasma with $B_0\approx 10$ kT, namely
\begin{eqnarray}
    a_0 = 1, \; b_0 = 1, \; \omega T = 15 \times 2\pi, \; \Gamma = 1,
\end{eqnarray}
one can obtain in case of the exact resonance at the end of the interaction process with the plane wave:
\begin{eqnarray}
    C_0 \left(\frac{b_0}{\Gamma} - 1\right) = 0.25, \; \omega_+ \approx 1.13, \; \omega_- \approx 0.87, \\
    p_x \sim \frac{1}{2 \Gamma} \left(C_0 \left(\frac{b_0}{\Gamma} - 1\right) \right)^2 (\omega T)^2 \approx 280, \\
    |p_y| \sim |p_z| \sim \left(C_0 \left(\frac{b_0}{\Gamma} - 1\right) \right) (\omega T) \approx 24,
\end{eqnarray}
which in dimensional units means electron energies of the sub-GeV range.
In experiment, however, these optimistic values can not be achieved for the whole plasma electrons, primarily because of the geometrical constraints. Moreover, the refractive index in plasma may violate the autoresonance condition \cite{kolom}, and collective plasma effects would add additional nonlinearities to the interaction and destroy the resonance at some moment. 
For these parameters, the radiation reaction effect is yet unimportant. According to the estimate \eqref{ratio}, for the parameters mentioned above one obtains
\beq
    \frac{F_{rr}}{F_L} \sim 10^{-4},
\eeq
so the radiation damping effects may be neglected compared with the Lorentz force in the considered ALA process.

For the resonant magnetic field shift in case of the Gaussian magnetic field profile, using \eqref{magn_param}, for the inhomogeneity spatial scale $\sigma=5\mu$m, one can obtain the value of the dimensionless amplitude $b_0$, using the characteristic transverse scale of electron trajectories. It can be found with integration of the equation of motion for the corresponding coordinate (e.g. $z$) considering the leading term in the transverse momentum to be proportional to $\xi \sin \xi$. Thus for $\Gamma = 1$, $\delta z\propto\xi \cos \xi$ and for the estimate of the l integration limits can be taken as $\omega T$ and $\omega T - \frac{\pi}{2}$. Then the estimate for the transverse oscillation amplitude is $|\delta z|_{max} \sim |p_z| \sim 24$ in dimensionless variables for the considered parameters. In CGS units this corresponds to $|\delta z|_{max} \sim~ 4 \mu$m. In phase average, $<|\delta z|> \sim \frac{|p_z|}{2} \sim 2~ \mu$m. An appropriate estimate for the dimensionless amplitude $b_0$ may be obtained for the value of $|\delta z|$ in between its maximum and average values, $b_0 \approx 1.08$.

\begin{figure*}[ht]
    \centering
    \includegraphics{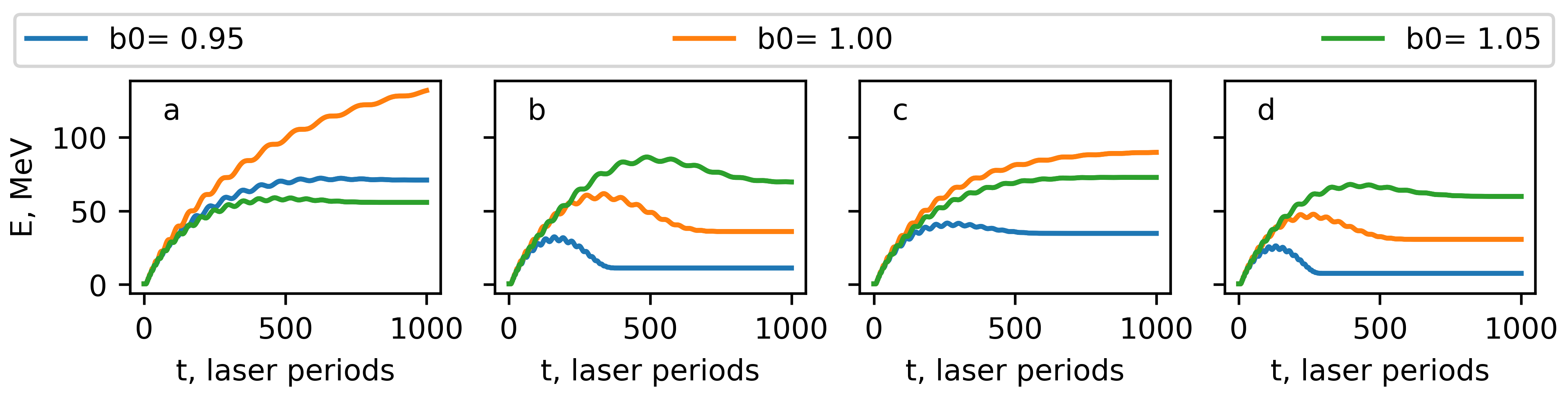}
    \caption{Numerical results for the energies of accelerated electrons for three different magnetic field amplitudes $b_0=\{0.95,~1.0,~1.05\}$; in all cases the incident wave has "$\sin^2$" temporal envelope; the applied magnetic field is directed along the axis of the laser pulse propagation direction. Panel a: the plane wave and the uniform magnetic field. Due to the autoresonance, which is met at $b_0=1.00$, it's continuously gaining energy from the pulse and is accelerated more effectively compared to other cases with $b_0 \neq 1.00$.
    Panel b: the plane wave and the magnetic field with the Gaussian transverse profile \eqref{magn_field}. Energy gain is sufficiently lower than in case of the uniform magnetic field and the resonance magnetic field amplitude is shifted to higher values. Panel c: The Gaussian pulse and the uniform magnetic field. Initially immobile electron placed at the focus on the laser axis is irradiated by the pulse with the beam waist radius of $10~ \mu$m. Final energy is sufficiently lower comparing to the case of the plane wave, however the electron is still ultrarelativistic. Panel d: The Gaussian pulse and the magnetic field with the Gaussian transverse profile \eqref{magn_field}. As in the case of the plane wave with inhomogeneus magnetic field, the energy is lowered and the resonance magnetic field amplitude is shifted to higher values.}
    \label{fig:energy}
\end{figure*}

\section{Effects of experimental constraints on autoresonant acceleration}\label{numerics}

In this section, numerical modeling for the autoresonant acceleration process are presented for sequentially more and more complicated situations, namely for single electron with optimal position in Section \ref{Single-particle model}, ensemble of non-interaction electrons with initial spatial distribution in Section \ref{Ensemble of non-interacting electrons} and 2D plasma system with the collective interactions taken into account in Section \ref{Plasma system with collective effects}.

\subsection{Single electron: laser and magnetic field inhomogeneities\label{Single-particle model}}

Consider a plane wave defined according to Eqs.\eqref{E}, \eqref{B}, with the wavelength $\lambda = 1 \mu $m, the pulse duration $T = 15 ~\lambda/c \approx 50~${fs},  $a_0 = 1$, which means the intensity of the order of $10^{18}~{\text{W}}/{\text{cm}^2}$.
To analyze electron acceleration process, equations of electron motion were solved with the Runge-Kutta method. 

Numerical simulation results for this situation show a good agreement with the analytical predictions. When an electron appears in the relativistic laser pulse and in a uniform magnetic field directed along the laser propagation direction, acceleration regime can be achieved with an appropriate external field amplitude. For the initially immobile electron the autoresonance is met at $b_0=1.00$, see the orange curve in Fig.~\ref{fig:energy}a; the final electron energy corresponds to the estimate \eqref{scalings_plane_wave}. Electron energy gain is maximal for the $b_0=1.00$ case, while 5\% deviation from this resonant value provides much smaller final energies, though still ultrarelativistic, which corresponds to near-resonant acceleration regime. 

 \begin{figure}[ht]
     \centering
     \includegraphics{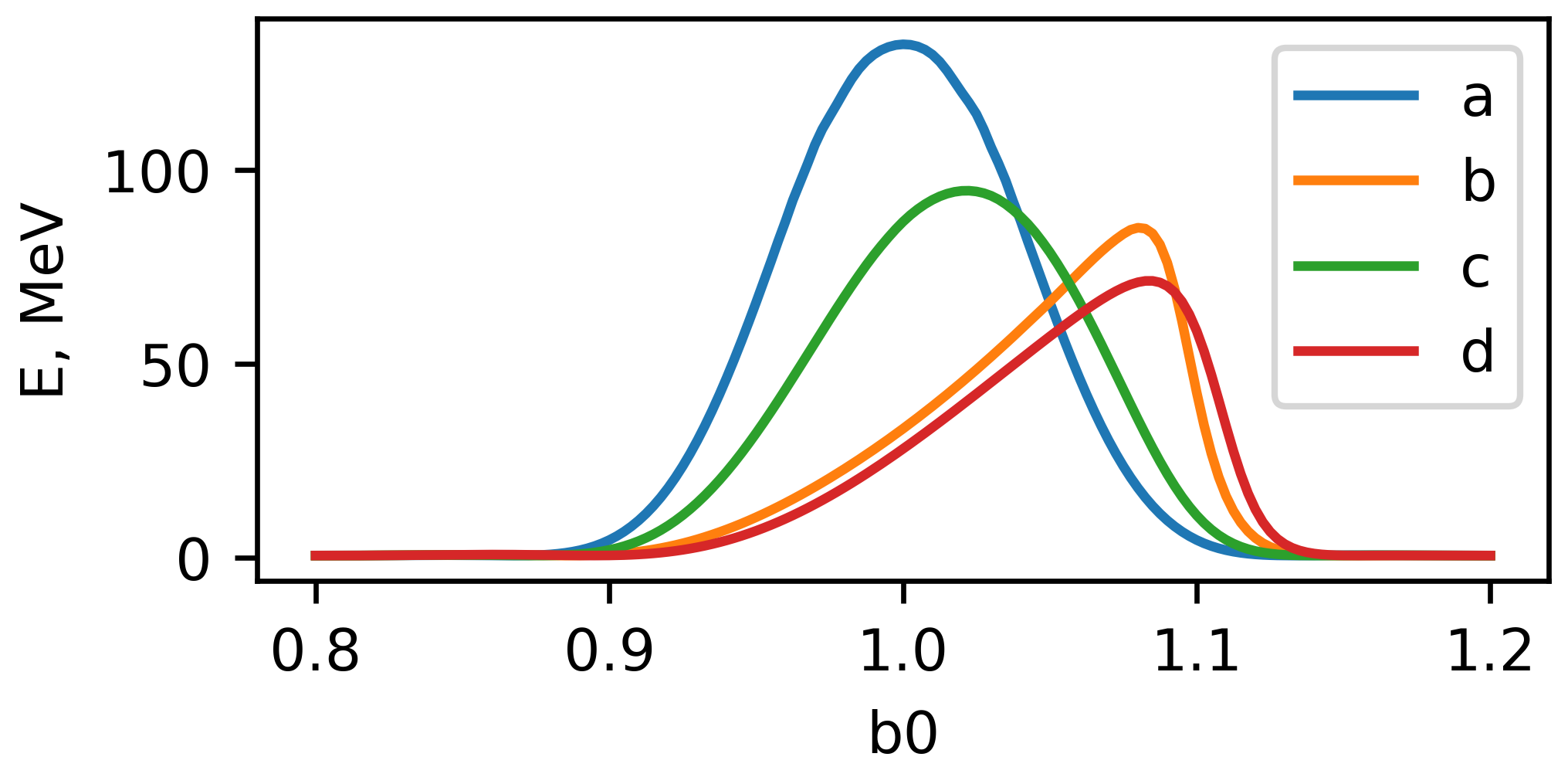}
     \caption{Electron energy gain as a function of the applied magnetic field amplitude. Curve a: the plane wave and the uniform magnetic field, curve b: the plane wave and the magnetic field with a Gaussian transverse profile, curve c: the Gaussian laser pulse and the uniform magnetic field, curve d: the Gaussian laser pulse and the magnetic field with the Gaussian transverse profile. Small shift of the near-resonant magnetic field amplitude ($b_0=1$ for the plane wave and the uniform magnetic field), is clearly visible.}
     \label{fig:B_data}
 \end{figure}

Considering the geometrical effect of the magnetic field profile, set the field according to \eqref{magn_field} with the characteristic spatial scale for the magnetic field inhomogeneity $\sigma=5~\mu$m.
Results for this case are shown in Fig.~\ref{fig:energy}b. The greatest energy gain is obtained for $b_0=1.05$, so the amplitude of the resonant magnetic field is increased. Transverse profile of the external magnetic field plays an important role in the electron acceleration process lowering the energy gain and changing the resonant magnetic field value.

A similar lowering of the final electron energy with a less resonant magnetic field shift is seen in Fig.~\ref{fig:energy}c, which corresponds to the case of an initially immobile electron placed on the laser axis at the focus for the Gaussian pulse (\ref{gaussE}, \ref{gaussB}) with $w_0=10~\mu$m and the uniform magnetic field. In this case, the magnetic field inhomogeneity also leads to the resonant amplitude field shift, see Fig.~\ref{fig:energy}d, thus causing a small final electron energy lowering due to the near-resonant acceleration regime.

The resonant magnetic field amplitudes for different situations are shown in Fig.~\ref{fig:B_data}. As mentioned above there is a shift of the resonant value due to the Gaussian transverse laser and magnetic fields profiles. The greatest effect takes the magnetic field inhomogeneity according to the close resonant peaks location for the cases of the Gaussian laser pulse profile and the magnetic field with uniform (curve c) and the Gaussian transverse (curve d) profiles.
Though both the laser and the magnetic field inhomogeneities sufficiently lower the electron energy gain and affect the resonance, which is automatically maintained only in the plane wave and the uniform magnetic field, nevertheless there is an effective acceleration regime for an appropriately adjusted magnetic field value that produces ultrarelativistic electrons as seen in Figs. ~\ref{fig:energy}, \ref{fig:B_data}.

\subsection{Ensemble of non-interacting electrons in laser and magnetic field inhomogeneities\label{Ensemble of non-interacting electrons}}

For electrons in an ensemble, the resonance conditions may be different. Dependence of the gained energy on initial electron position and momentum was investigated with the Monte-Carlo method. Simulations were performed for the same set of parameters ($\lambda = 1 \mu $m, $T = 15 \lambda/c$,  $a_0 = 1$, $\sigma=5~\mu$m). Statistics involve 1000 particles in each case. 
Initial spatial distribution of particles was uniform within the box $2 w_0\times2 w_0\times 2 w_0$, initial momentum distribution was Maxwellian with the temperature $10$ keV. The box center is located at the intersection of the focal plane by the laser axis for the Gaussian beam. In the plane wave with a uniform magnetic field the box position has no effect on electron spectrum, in other cases it appears to play an important role in electron dynamics. 

\begin{figure}[h]
    \includegraphics[width=\linewidth]{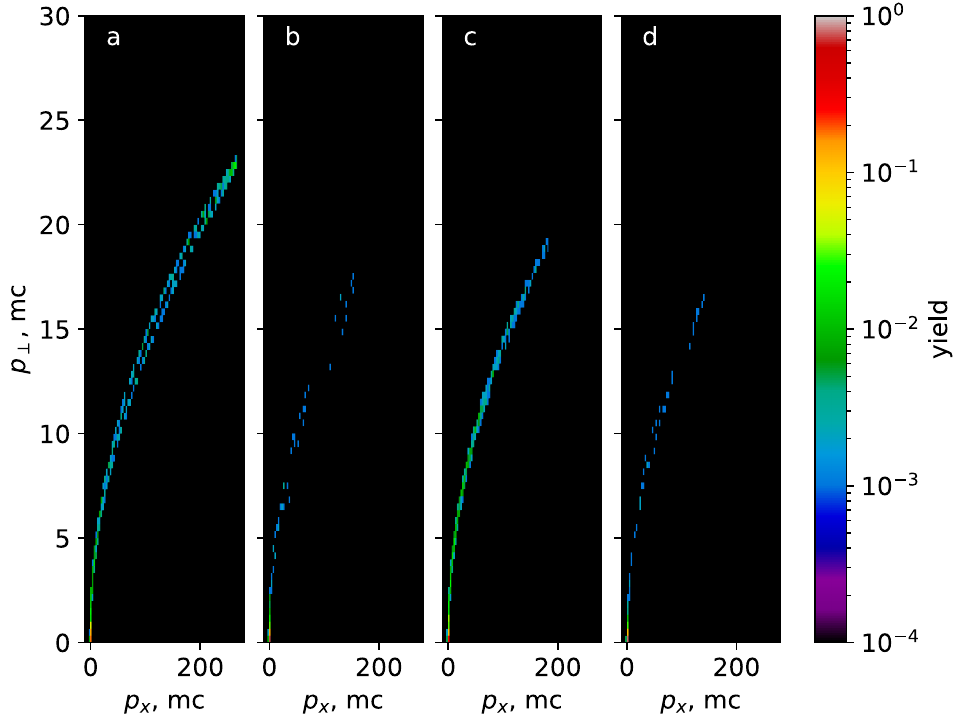}
    \caption{Momentum phase space for 1000 particles with initially uniform spatial and Maxwellian momentum distributions. Panel a: the plane wave and the uniform magnetic field with $b_0=1.00$; panel b: the plane wave with the Gaussian magnetic field with $b_0=1.08$; panel c: the Gaussian pulse and the uniform magnetic field with $b_0=1.02$; panel d: the Gaussian pulse and the Gaussian magnetic field with $b_0=1.08$. The most effective configuration is the "a", the others result in a less number of energetic electrons and a less value of their maximal longitudinal momentum.}
    \label{fig:monte}
\end{figure}

Simulations results are shown in Fig.~\ref{fig:monte}, which represents the momentum space distribution after 1000 laser periods. The most effective acceleration with the highest value of the longitudinal momentum $p_x$ and the greatest number of energetic particles is seen when they are irradiated by the plane wave in the uniform magnetic field with $b_0=1.00$, see Fig.~\ref{fig:monte}a. Fig.~\ref{fig:monte}b corresponds to the plane wave and the Gaussian magnetic field with $b_0=1.08$. The total number of energetic particles and the maximal longitudinal momentum is sufficiently lower because the resonance condition breaks for a substantial number of electrons. Fig.~\ref{fig:monte}c and Fig.~\ref{fig:monte}d represent the momentum space for electrons irradiated by the Gaussian pulse with $w_0=10~\mu$m in the uniform ($b_0=1.02$) and the Gaussian ($b_0=1.08$) magnetic fields. The trend for an effectiveness lowering in the magnetic field with the transverse Gaussian profile is seen similarly to the previous single-particle results in Fig.\ref{fig:energy}. Slow phase matching breaking due to the inhomogeneous profile of the laser field additionally lowers the number of accelerated electrons number and their longitudinal momentum.

\subsection{Autoresonant acceleration of electrons in finite magnetized plasmoid\label{Plasma system with collective effects}
}

For analysis of plasma effects in the autoresonant acceleration process, a set of 2D Particle-in-cell simulations were performed with the open-source code Smilei \cite{smilei} for the same set of laser parameters ($\lambda = 1 \mu $m, $T = 15 \lambda/c \approx 50~${fs}, $a_0 = 1$). Several values of plasma density were analyzed. All electrons are initially immobile and in the simulations of the first type are distributed uniformly within a plane of 10 $\mu$m in the transverse direction. It is Gaussian along the laser propagation direction with a characteristic width $\approx$ 11 $\mu$m and is cut at the level of $10^{-6}$ of the maximum value. The plasma layer is located at $x=15~\mu$m. In these simulations the magnetic field was uniform $b_0 = 1.00$.

It appeared that to obtain a good agreement with the analytical benchmarks, in numerical simulations the resolution should be sufficiently high, at least for the resonant trajectories, it was chosen as 100 cells per wavelength in the longitudinal direction and 5 cells per wavelength in the transverse one.

\begin{figure}[h]
    \includegraphics[width=\linewidth]{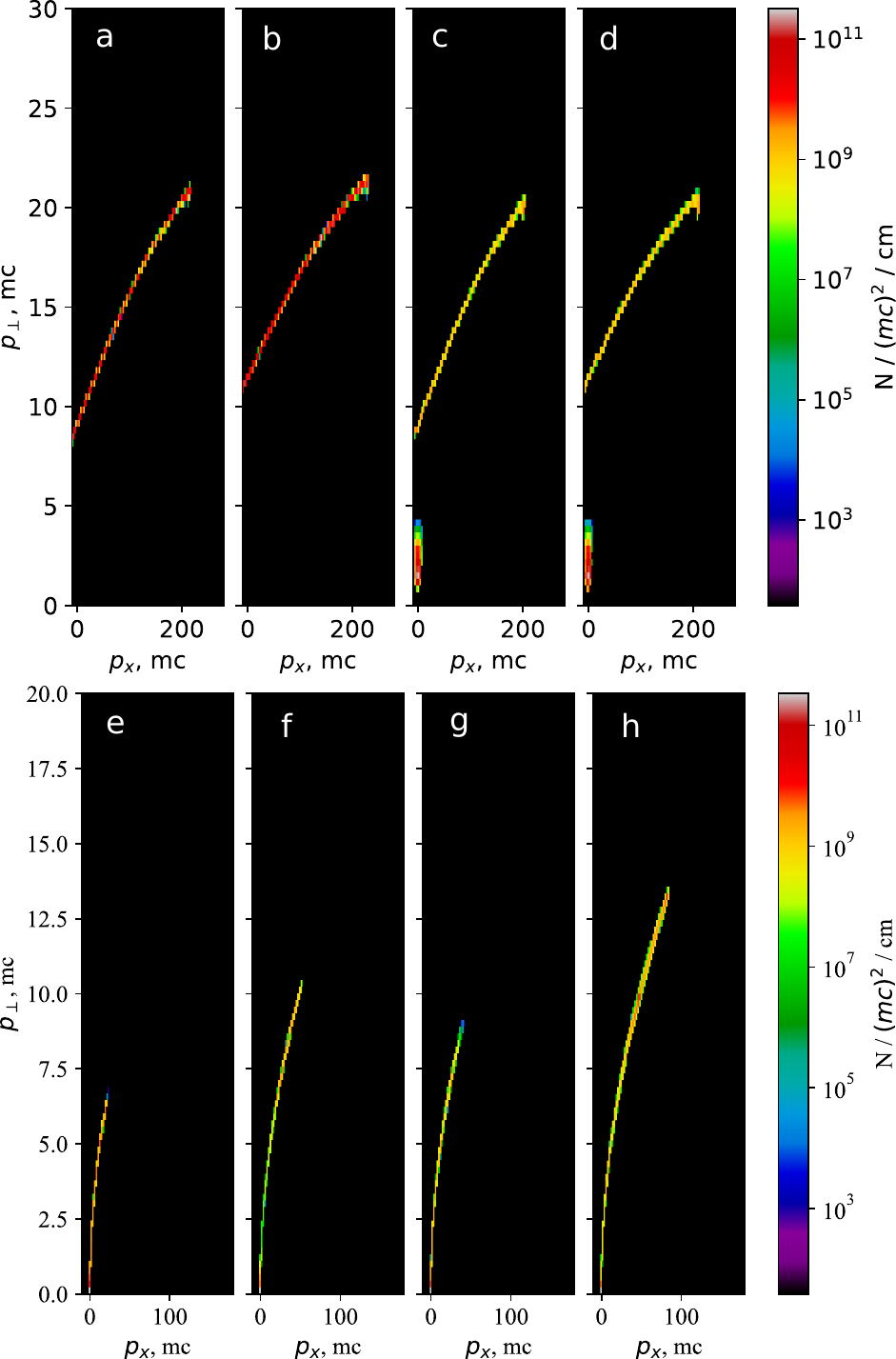}
    \caption{Panels a-d: results of 2D PIC simulations for a plasma layer and a uniform magnetic field. Momentum space for low ($10^{-3} \; N_{cr}$) and high ($10^{-2} \; N_{cr}$) densities for two moments of time: after $800$ and $1600$ laser periods. Panel a: low density case after 800 laser periods; panel b: low density case after 1600 laser periods; panel c: high density case after 800 laser periods; panel d: high density case after 1600 laser periods.\\
    Panels e-h: results of 2D PIC simulations for a plasma cylinder for two laser intensities and two moments of time. Panel e: momentum space for $a_0 = 1$ after 45 laser periods; panel f: momentum space for $a_0 = 1$ after 90 laser periods; panel g: momentum space for $a_0 = \sqrt{10}$ after 45 laser periods; panel h: momentum space for $a_0 = \sqrt{10}$ after 90 laser periods;
    }
    \label{fig:density+cylinder}
\end{figure}

In case of dilute plasma, the results obtained in frames of the single particle model are reproduced with a high accuracy, which may be attributed to the insignificance of collective effects in this situation. Increasing the plasma density makes the electron momentum distribution wider and at the same time increase the total charge of the accelerated beam. There is an optimum, when the momentum spread is not too great and the number of accelerated electrons is not too small, which is around $10^{-2}...10^{-3}N_{cr}$, where $N_{cr}$ is the plasma critical density, see Fig.~\ref{fig:density+cylinder}a and Fig.~\ref{fig:density+cylinder}b. For the higher density $10^{-2}N_{cr}$, there is a deceleration for a sufficient part of electrons, observed in the electron spectrum in Fig.~\ref{fig:density+cylinder}c and Fig.~\ref{fig:density+cylinder}d.

\begin{figure}[ht]
    \centering
    \includegraphics[width=\linewidth]{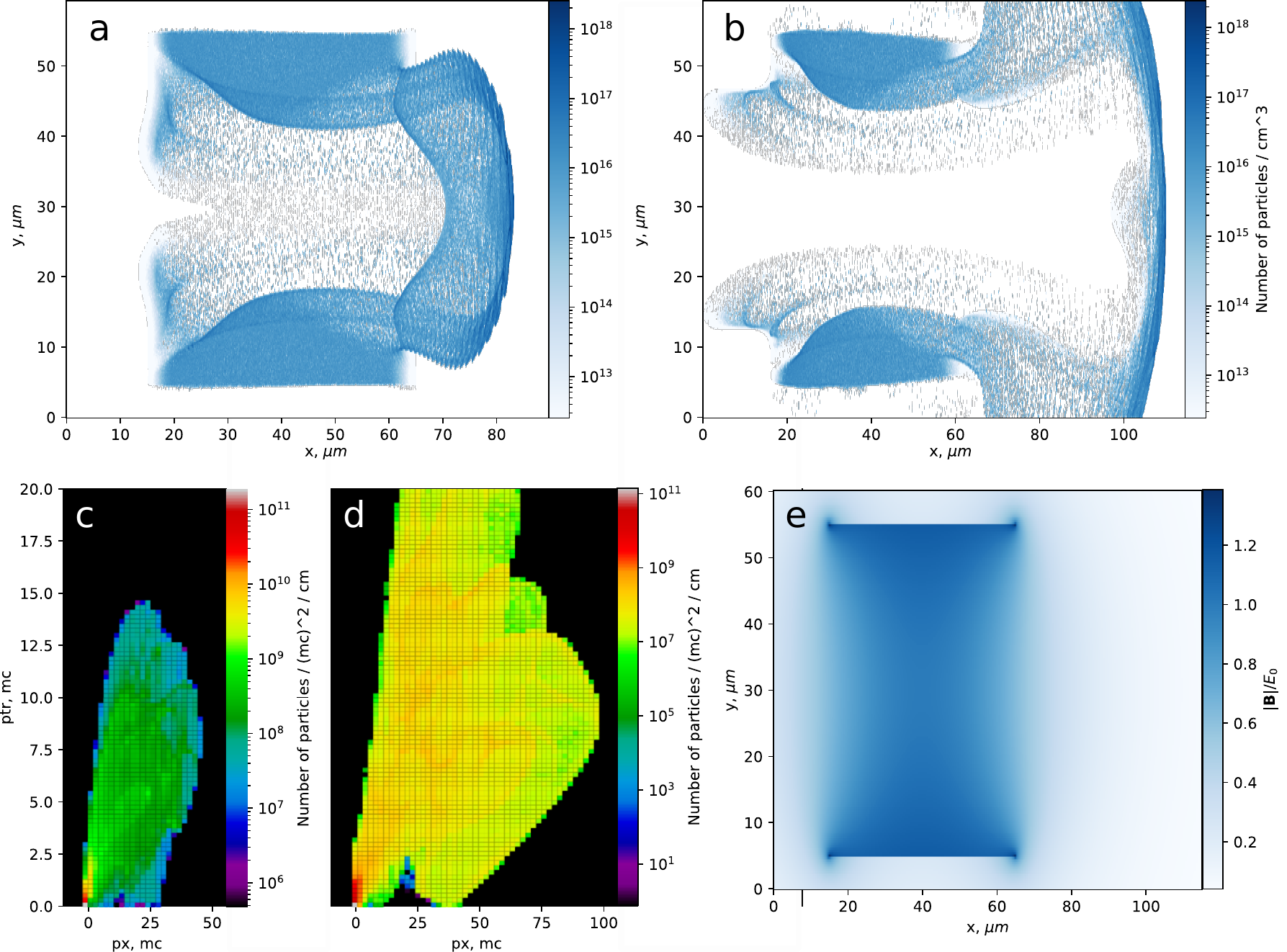}
    \caption{Electron acceleration in a cylinder with solenoid magnetic field applied for two Gaussian laser pulse intensities after 115 laser periods. Panels a and b represent electron density distribution for $a_0 = 1$ and $a_0 = \sqrt{10}$ respectively. Close to the cylinder edges electrons become bent following magnetic field lines. With increased intensity they are suppressed in a longitudinal direction and spread in a transverse one as seen on panel b. Panels c and d represent electron momentum space distribution for the same $a_0$. Compared to the one in gaussian magnetic field this distribution is much wider. Magnetic field in the simulation is shown on panel e. It is inhomogeneous in both, transverse and longitudinal directions, especially close to cylinder boundaries.}
    \label{fig:solenoid}
\end{figure}

In a more realistic situation, the plasma occupies a much larger area. Increasing the plasma length may be beneficial for accelerating more particles but also increases the charge-separation electric field, which affects the autoresonance. In the simulations of the second type, the longer plasma target and also the spatial distribution of the laser beam and the magnetic field profile were introduced. The parameters of the magnetic field were optimally adjusted to an efficient acceleration of a greater number of electrons. Initially cold electrons are distributed within a cylinder with radius of $ 25 \; \mu$m and length of $50 \; \mu$m. Plasma is uniform in the transverse direction and has the following longitudinal profile: it is uniform except Gaussian decay on both sides on $5 \; \mu$m scale. Particle density in the plasma cylinder is $10^{-5} \; N_{cr}$. Magnetic field is Gaussian with $\sigma = 10 \; \mu$m and $b_0=1.08$. Two intensities of the laser driver were considered, $a_0 = 1$ and $a_0 = \sqrt{10}$.

Momentum space distributions of the electrons irradiated by the pulse with $a_0 = 1$ after $45$ and $90$ laser periods are shown in Fig.~\ref{fig:density+cylinder}e and Fig.~\ref{fig:density+cylinder}f respectively. Due to the laser and the external magnetic field inhomogeneities, electron acceleration rate and the ratio between longitudinal and transverse momenta are lower than in the plane wave, see Fig.~\ref{fig:density+cylinder}a and Fig.~\ref{fig:density+cylinder}b. However, for the optimally adjusted magnetic field strength and appropriate plasma density and size, generation of an ultrarelativistic beam is possible. Lasers with higher intensities provide more effective acceleration as shown in Fig.~\ref{fig:density+cylinder}g and Fig.~\ref{fig:density+cylinder}h. Note that according to the analytical model for the plane wave, the longitudinal momentum of the fastest particles \eqref{exactp3} scales linearly with the laser intensity, but the results presented in Fig.~\ref{fig:density+cylinder}g and Fig.~\ref{fig:density+cylinder}h show a worse improvement, which can be explained by the plasma collective and field inhomogeneity effects.

Finally, the third type of the PIC simulations was performed, where the magnetic field profile corresponds to the magnetic field in a plasma solenoid, which generally corresponds to the optically driven strongly magnetized plasmoids \cite{Nuter.etal_PlasmaSolenoidDriven_PRE-2018, chiral}. In this situation, in addition to the geometrical effects, considered in the previous simulation setup, the edge effects of the magnetic field profile are added. The plasma cylinder with the parameters defined above, put in a solenoidal magnetic field, is irradiated with the same Gaussian laser pulses. 
The magnetic field profile is seen in Fig.\ref{fig:solenoid}e. Close to the plasma edges the field is highly inhomogeneous and rapidly decreases outside along the axes. Magnetic field amplitude is $1.08$ at the center of the cylinder.

Electron density distribution after $115$ laser periods is shown in Fig.~\ref{fig:solenoid}a for $a_0 = 1$ and in Fig.~\ref{fig:solenoid}b for $a_0 = \sqrt{10}$. It is seen that the laser pulse has pushed most of the cylinder electrons located close to the axis in the propagation direction. 
In this case, due to the complicated magnetic field profile, electrons accelerate even less effectively, having a wider distribution in the momentum space, see Fig.~\ref{fig:solenoid}c and Fig.~\ref{fig:solenoid}d.
Actually, the momentum distribution widens dramatically at the back side of the solenoid, where the magnetized electron flow follows the magnetic field lines for some time before getting free.
The observed spread in the momentum phase space makes this most realistic configuration the least efficient in terms of the autoresonant acceleration of electrons up to ultrarelativistic energies, but it still looks promising, as it allows to obtain a relatively high number of sub-GeV electrons with conventional relativistic laser setups.   

\section{Conclusions}

The presented analysis shows that the autoresonant acceleration of electrons is quite a stable process, which survives under small modifications for spatially inhomogeneous laser and magnetic fields, and with accounting for plasma collective effects. In a real setup, which may be performed as an optically driven strongly magnetized plasma solenoid, the most important effect on the acceleration process is related to the fast magnetic field decay at the rare side of the plasma. The magnetic field profile of the magnetized plasma solenoid may be in principle controlled by the initial target properties, however, even the considered non-optimized case demonstrates the high number of ultrarelativistic electrons and their relatively small angular spread. The obtained results proposes in general a possibility to adjust the parameters of the irradiated micro-plasmoid and the laser driver for an effective all-optical electron acceleration scheme.

\section*{Acknowledgements}
The work was supported by the Foundation for the Advancement of Theoretical Physics and Mathematics “BASIS”. We are grateful to the High-Performance Computing Center of the National Research Nuclear University MEPhI and the Joint Supercomputer Center of the Russian Academy of Sciences for the provided resources.


%

\end{document}